\documentclass[10pt, twoside, twocolumn]{IEEEtran}
 \makeatletter

\newcommand{\Rmnum}[1]{\expandafter\@slowromancap\romannumeral #1@}
\makeatother
\usepackage{graphicx,cite,epsfig,amssymb,amsmath,multirow}
\usepackage{graphicx,amsmath,amssymb,amsfonts,cite}
\usepackage{mathrsfs}
\usepackage{algorithm}
\usepackage{algorithmic}
\usepackage{float}
\usepackage{bm}
\usepackage{booktabs}
\usepackage{epstopdf}
\usepackage{ulem}
\usepackage{color}
\usepackage{makecell}
\newcommand{\tabincell}[2]{\begin{tabular}{@{}#1@{}}#2\end{tabular}}

\makeatother

\begin{document}
\bibliographystyle{ieeetr}

\title{FDM-Structured Preamble Optimization for Channel Estimation in MIMO-OQAM/FBMC Systems}

\author{Wenfeng~Liu,~Da~Chen,~Kai~Luo,~Tao~Jiang,~and~Daiming~Qu
\thanks{W.~Liu,~D.~Chen,~K.~Luo,~T.~Jiang,~and~D.~Qu are with School of Electronic Information and Communications, Wuhan National Laboratory for Optoelectronics, Huazhong University of Science and Technology, Wuhan 430074, China (e-mail: liuwenfeng@hust.edu.cn; chenda@hust.edu.cn; kluo@hust.edu.cn; tao.jiang@ieee.org; qudaiming@hust.edu.cn).}}


\maketitle

\IEEEpubid{This work has been submitted to the IEEE for possible publication. Copyright may be transferred without notice, after which this version may no longer be accessible.}

\begin{abstract}
In this paper, we consider the problem of preamble design in multiple-input multiple-output (MIMO) systems employing offset quadrature amplitude modulation based filter bank multicarrier (OQAM/FBMC) and propose a preamble optimization method for the frequency division multiplexing (FDM)-structured preamble. Specifically, we formulate an optimization problem to determine the frequency division multiplexed preambles, where the objective is to minimize the mean square error (MSE) of the channel estimation, subject to the constraint on the transmit energy. For two transmit antennas, we find the relationship between preambles and the intrinsic interference from neighboring symbols to achieve the minimum channel estimation MSE, and derive the optimal closed-form solution. For more than two transmit antennas, the constrained preamble optimization problem is nonconvex quadratic. Therefore, we convert the original optimization problem into a quadratically constrained quadratic program (QCQP) and obtain the suboptimal solution by relaxing the nonconvex constraint. Simulation results demonstrate that, in terms of MSE and bit error rate (BER) performances, the proposed method outperforms the conventional FDM preamble design method at all signal-to-noise ratio (SNR) regimes and outperforms the interference approximation method-complex (IAM-C) preamble design method at low to medium SNR regimes with lower preamble overhead.
\end{abstract}


\IEEEpubidadjcol

\section{INTRODUCTION}
\IEEEPARstart{M}{ulticarrier modulations (MCM)} have attracted a lot of attention due to the capability to efficiently cope with frequency selective channels. Much of the attention in the present literature emphasizes on the use of the conventional orthogonal frequency division multiplexing (OFDM). However, the OFDM system uses rectangular window on each subchannel, which leads to high out-of-band radiation. Moreover, the OFDM system sacrifices data transmission rate because of the insertion of cyclic prefix (CP). To remedy the problems of the OFDM system, the offset quadrature amplitude modulation based filter bank multicarrier (OQAM/FBMC) has attracted increasing attention \cite{kd2014, Farhang2011, daiqu, gao2002, lei2017, xiqigao, doubly2017}. Compared with the conventional OFDM system, the OQAM/FBMC system provides lower sidelobes through  the use of well-shaped prototype filter and higher useful data rate due to the fact that OQAM/FBMC does not require the CP. Furthermore, OQAM/FBMC brings advantages such as robustness to narrow-band interference and carrier frequency offset. Due to the above superiorities over OFDM, OQAM/FBMC is being considered as a promising technique for cognitive radio \cite{PHYsical}, professional mobile radio (PMR) evolution \cite{EMPhAtiC}, and 5G cellular networks \cite{Banelli2014, QLi2014}. As another promising technique in future communication systems, multiple-input multiple-output (MIMO) is able to increase the system throughput and the link reliability \cite{Gesbert2002, SJin}. Therefore, it is believed that the successful combination of MIMO and OQAM/FBMC is important and is envisioned to achieve higher spectral efficiency \cite{P¨¦rez, Rottenberg2017, Mestre2016}.

In MIMO-OQAM/FBMC communication systems, the channel estimation plays a significant role in the recovery of data symbols. Typically, training symbols are employed at the transmitter to perform channel estimation. In the literature, several training schemes and associated estimation methods for MIMO-OQAM/FBMC systems have been proposed, which can be mainly divided into two categories, i.e., the scattered pilots-based methods \cite{rupp2016, Javaudin, Javaudin2008} and the preamble-based methods \cite{Kofidis2013,Kofidis2011, KofidisEW, jar2009, Taheri2015}. Generally speaking, the scattered pilots-based methods are used to track the channel variations in fast fading environments, while the preamble-based methods are more suitable for time invariant channels with better channel estimation performance. In this paper, we consider the time invariant channels with low frequency selectivity and focus on the preamble-based methods in MIMO-OQAM/FBMC systems.

The appropriate preamble design is crucial for channel estimation in MIMO-OQAM/FBMC systems. Different from the OFDM system, the orthogonality condition of the OQAM/FBMC system only holds in the real field, which causes intrinsic imaginary interference between data symbols and preamble symbols \cite{dachen, Saeedi2011}. Note that, the OFDM system provides orthogonality in the complex field and there is no imaginary interference between symbols. Therefore, the conventional preamble design methods in the OFDM system cannot be directly used in the OQAM/FBMC system \cite{wenfeng, kd}. Moreover, when OQAM/FBMC is combined with MIMO techniques, the multi-antenna interference should also be considered, which makes the preamble design more complicated.

\IEEEpubidadjcol

The typical preamble design method, interference approximation method (IAM), has been proposed and widely investigated in MIMO-OQAM/FBMC systems due to its simplicity and efficiency \cite{Kofidis2013, Kofidis2011}. The key idea is to approximate the intrinsic interference from neighboring symbols and construct complex-valued \emph{pseudo-pilots} at all frequencies to estimate the channel. However, since the preamble overhead per antenna increases linearly with the number of transmit antennas, the IAM method encounters a considerable loss in spectral efficiency, which is unbearable especially for a relatively large number of transmit antennas. To reduce the preamble overhead, some other effective methods with shorter preambles were proposed in MIMO-OQAM/FBMC systems, which include the time domain channel model-based preamble design method \cite{KofidisEW} and the sparse preamble design methods with only three OQAM/FBMC symbols being needed per antenna for channel estimation \cite{jar2009, Taheri2015}. The frequency-division multiplexing (FDM) based preamble design was first introduced in \cite{HMinn2006} for MIMO-OFDM systems and this idea reappeared recently in \cite{jar2009, Taheri2015} in the context of MIMO-OQAM/FBMC systems, which shows great channel estimation performance with low preamble overhead. In the FDM method, the preamble symbols from different transmit antennas are frequency division multiplexed. As a consequence, the channels from each of the transmit antennas can be estimated separately without inter-antenna interference. Compared with the IAM method, the FDM method requires only three columns of preamble overhead and hence achieves a significant improvement in spectral efficiency. However, the preamble in the existing FDM method is obtained without theoretical analysis and its optimality cannot be guaranteed, resulting in a significant performance loss in channel estimation. Therefore, the preamble optimization for the FDM-structured preamble in MIMO-OQAM/FBMC systems needs further investigations.

This paper aims to design the optimal FDM-structured preamble sequence to enhance the estimation accuracy. The key idea is to formulate a preamble optimization problem by minimizing the mean square error (MSE) of the channel estimation, with the constraint on the training energy at the MIMO-OQAM/FBMC transmitter. In the case of two transmit antennas, we derive the closed-form solution for the preamble optimization problem by exploring the relationship between preambles and the intrinsic interference from neighboring symbols to achieve the minimum MSE. In the case of three or more transmit antennas, the preamble optimization problem is nonconvex quadratic, which causes the optimal solution hard to obtain. Therefore, we convert the original optimization problem into a quadratically constrained quadratic program (QCQP) and relax the nonconvex constraint of the optimization problem to obtain the suboptimal solution. In order to verify the validity of the proposed preamble optimization method, and compare the MSE and bit error rate (BER) performances with the conventional FDM and IAM methods, simulations are conducted under $2\times2$ and $4\times4$ MIMO configurations, respectively. Simulation results show that the proposed method achieves better performance than the conventional FDM method at all signal-to-noise ratio (SNR) regimes and outperforms the IAM-real (IAM-R) and IAM-complex (IAM-C) methods (two variants of IAM) at low to medium SNR regimes with lower preamble overhead.

The remainder of the paper is organized as follows. In Section II, the MIMO-OQAM/FBMC system model is presented. The conventional IAM and FDM preamble design methods are reviewed in Section III. Then, the preamble optimization problem for the FDM-structured preamble and the different solving processes that correspond to different number of transmit antennas are proposed in Section IV. Section V compares the MSE and BER performances of different preamble design methods through simulations, followed by the conclusions in Section VI.

The following notations are employed in this paper. Bold lower-case letters denote column vectors and bold upper-case letters are used for matrices. For the matrix $\textbf{A}$, notations $\textbf{A}^T$, $\textbf{A}^H$ and $\textrm{Tr}(\textbf{A})$ indicate its transpose, conjugate transpose and trace operations, respectively. $\Re\{\cdot\}$ is the real part of a complex-valued number and $\mathbb{E}(\cdot)$ is the expectation operator. We use $\textbf{I}_{n}$ and $\textbf{0}_{n}$ to denote an $n\times n$ identity matrix and a zero matrix, respectively. The notation $\mathbb{C}^{m\times n}$ represents a set of $m\times n$ matrices with complex entries. Finally, $j=\sqrt{-1}$.

\section{SYSTEM MODEL}
\subsection{OQAM/FBMC System Model}
The baseband discrete-time signal at the output of an OQAM/FBMC synthesis filter bank (SFB) can be written as \cite{Siohan2002, CE}
\begin{equation}\label{eq 2.01}
{s[k]}=\sum_{m=0}^{M-1}\sum_{n\in{\mathbb{Z}}}a_{m,n}\underbrace{g\left [k-n\frac{M}{2}\right ]e^{j2\pi mk/M}e^{j\varphi_{m,n}}}_{g_{m,n}[k]},
\end{equation}
where $M$ is the number of subcarriers, $a_{m,n}$ denotes the real-valued symbol conveyed by the subcarrier of index $m$ during the symbol time of index $n$, and $g[k]$ is a symmetrical real-valued prototype filter satisfying the perfect reconstruction condition \cite{Siohan2002}. $\varphi_{m,n}$ is an additional phase term given by $\varphi_{m,n}=\varphi_{0}+(\pi/2)(m+n)$ mod $\pi$, where $\varphi_{0}$ can be arbitrarily chosen. Here we set $\varphi_{0}=0$.

We assume that the OQAM/FBMC signal is transmitted through a channel that varies slowly with time, and its delay spread is significantly shorter than the symbol interval \cite{doubly2017}. This implies that the channel transfer function over each subcarrier band may be approximated by a flat gain. Let $H_{m,n}$ denotes this gain for the $m$th subcarrier at the $n$th symbol interval. With the slow varying assumption, we omit the subscript $n$ from $H_{m,n}$ for the sake of brevity. In addition, a complex additive white Gaussian noise (AWGN) with zero mean and variance $\sigma^2$ is assumed to be introduced at the channel output. Accordingly, we can express the analysis filter bank (AFB) output at the $m$th subcarrier and $n$th OQAM/FBMC symbol as \cite{CE}
\begin{equation}\label{eq 2.02}
y_{m,n}=H_{m}c_{m,n} + \eta_{m,n},
\end{equation}
where $\eta_{m,n}$ originates from the channel noise and $c_{m,n}= a_{m,n} + ja_{m,n}^{(i)}$ is the virtually transmitted symbol at $(m,n)$, with $ja_{m,n}^{(i)}$ being the intrinsic imaginary interference from the neighboring frequency-time (FT) points. With a well localized pulse $g[k]$ in time and frequency, it can be assumed that the intrinsic imaginary interference mostly originates from the first-order neighboring  FT points \cite{CE}.
\begin{figure*}
\centering
\includegraphics[scale=0.85]{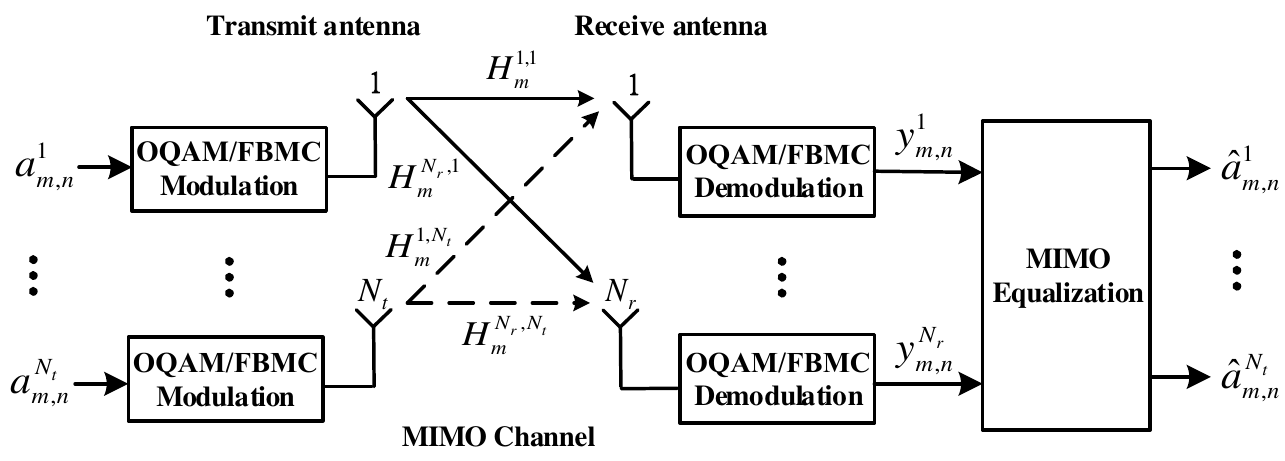}
\caption{The MIMO-OQAM/FBMC system model.}
\label{fig110}
\end{figure*}
Let us denote
\begin{equation}\label{eq 2.002}
\zeta_{m,n}^{p,q}=\sum_{k=-\infty}^{\infty}{g_{m,n}[k]g_{p,q}^{\ast}[k]}.
\end{equation}
Then, the imaginary interference term $ja_{m,n}^{(i)}$ can be approximated as
\begin{equation}\label{eq 2.03}
ja_{m,n}^{(i)} \approx \sum\limits_{(p_{0},q_{0})\in\Omega_{1}} a_{m + p_{0},n + q_{0}}\zeta_{m + p_{0},n + q_{0}}^{m,n},
\end{equation}
where the neighborhood $\Omega _{1} = \{(p_{0},q_{0})|\ p_{0}, q_{0}\in\{-1, 0, 1\}\ \textrm {and} \ (p_{0},q_{0})\neq{(0,0)}\}$ and $\zeta_{m + p_{0},n + q_{0}}^{m,n}$ represents the contribution of $a_{m + p_{0},n + q_{0}}$ to the imaginary interference $ja_{m,n}^{(i)}$. It is noteworthy that for a well-designed prototype filter $g[k]$, $\zeta_{m + p_{0},n + q_{0}}^{m,n}$ is pure imaginary for any $(p_{0},q_{0})\neq{(0,0)}$ and $\zeta_{m + p_{0},n + q_{0}}^{m,n}=1$ for $(p_{0},q_{0})={(0,0)}$. Thus, for $(p_{0},q_{0})\neq{(0,0)}$, in the remainder of this paper, we refer to the terms $\zeta_{m + p_{0},n + q_{0}}^{m,n}$ as the \emph{imaginary interference coefficients}.

Note that, the intrinsic imaginary interference could be removed by taking the real part after channel equalization. In this paper, since we mainly focus on the preamble design issues, the equalization and operation of taking real part are not discussed.

\subsection{MIMO-OQAM/FBMC System Model}
The MIMO-OQAM/FBMC system model is depicted in Fig. \ref{fig110}, where the transmitter and the receiver are equipped with $N_{t}$ and $N_{r}$ antennas, respectively. Since the channel estimation performance of MIMO-OQAM/FBMC systems would not differ in the case of channel coding, we only consider the uncoded scenario below.

At the transmitter side, the symbols spatially multiplexed on the $m$th subcarrier at the $n$th time index are denoted by $\textbf{a}_{m,n} = \left[{a_{m,n}^{1}},{a_{m,n}^{2}},\cdots,{a_{m,n}^{N_{t}}} \right]^T$, each element of which is transmitted at different antennas after the corresponding OQAM/FBMC modulation. At the receiver side, the link of each transmit and receive antenna pair is degraded by multipath fading and contaminated with AWGN. For each given FT position $(m,n)$, let $H_{m}^{r,t}$ be the frequency response of the channel between the $t$th transmit antenna and the $r$th receive antenna and $\eta_{m,n}^{r}$ be the noise component at the $r$th receive antenna. By assuming perfect time and frequency synchronization, the demodulated symbol of the $r$th receive antenna can be obtained by extending (\ref{eq 2.02}) to the MIMO case as \cite{P¨¦rez, Kofidis2013}
\begin{equation}\label{eq 2.8}
y_{m,n}^{r} =\sum_{t=1}^{N_{t}} H_{m}^{r,t}c_{m,n}^{t}+\eta_{m,n}^{r}, \  1\leq r\leq{N_{r}},
\end{equation}
where $c_{m,n}^{t}$ represents the corresponding virtually transmitted symbol at the $t$th transmit antenna. According to (\ref{eq 2.03}), $c_{m,n}^{t}$ can be written as
\begin{equation}\label{eq 2.9}
c_{m,n}^{t}=a_{m,n}^{t}+\sum\limits_{(p_{0},q_{0})\in\Omega_{1}} a_{m + p_{0},n + q_{0}}^{t}\zeta_{m + p_{0},n + q_{0}}^{m,n}, \  1\leq t\leq{N_{t}}.
\end{equation}
We denote the demodulated symbol vector by $\textbf{y}_{m,n} = \left[{y_{m,n}^{1}},{y_{m,n}^{2}},\cdots,{y_{m,n}^{N_{r}}} \right]^T$, the virtually transmitted vector by $\textbf{c}_{m,n} = \left[{c_{m,n}^{1}},{c_{m,n}^{2}},\cdots,{c_{m,n}^{N_{t}}} \right]^T$ and the additive noise vector by $\bm{\eta}_{m,n} = \left[{\eta_{m,n}^{1}},{\eta_{m,n}^{2}},\cdots,{\eta_{m,n}^{N_{r}}} \right]^T$. Thus, the equation (\ref{eq 2.8}) can be expressed as
\begin{equation}\label{eq 2.10}
\textbf{y}_{m,n} = \textbf{H}_{m}\textbf{c}_{m,n} + \bm{\eta}_{m,n},
\end{equation}
where
\begin{align}\label{eq 2.11}
 \textbf{H}_{m}=\left[ \begin{array}{cccc}
H_{m}^{1,1}   & H_{m}^{1,2}   & \cdots   & H_{m}^{1,N_{t}}          \\
H_{m}^{2,1}   & H_{m}^{2,2}   & \cdots   & H_{m}^{2,N_{t}}          \\
\vdots   & \vdots    & \ddots          & \vdots                        \\
H_{m}^{N_{r},1}   & H_{m}^{N_{r},2}   & \cdots   & H_{m}^{N_{r},N_{t}}          \\
\end{array} \right]
\end{align}
is the MIMO channel frequency response (CFR) at that FT point.

\section{THE CONVENTIONAL PREAMBLE DESIGN METHODS IN MIMO-OQAM/FBMC SYSTEMS}
Since the orthogonality condition of the OQAM/FBMC system only holds in the real field, which causes intrinsic imaginary interference to preamble symbols at the receiver, preamble design in the OQAM/FBMC system is more difficult than that in OFDM. Worse is that there also exists multi-antenna interference when OQAM/FBMC is extended to the MIMO case. In this section, we tackle the preamble design problem in MIMO-OQAM/FBMC systems and give a brief review of the conventional preamble design methods.

\subsection{IAM Method}
\begin{figure}[!t]
\centering
\includegraphics[scale=0.8]{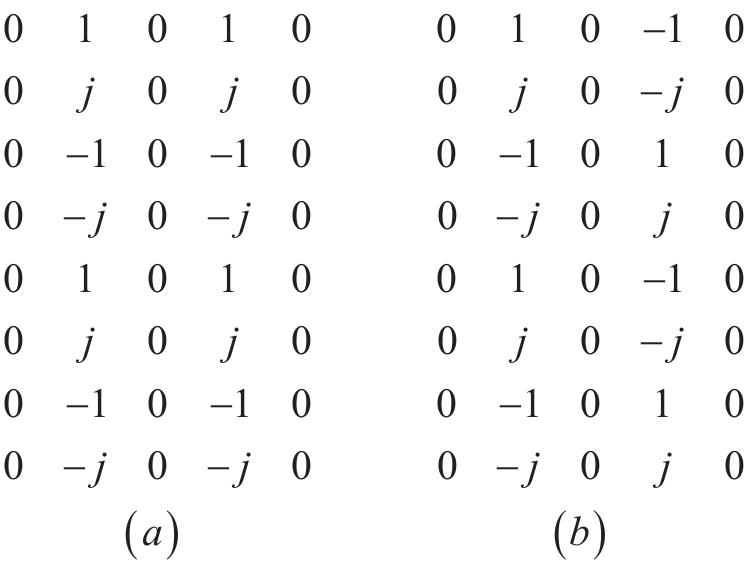}
\caption{Transmitted pilots at the SFB input for the IAM-C method, with (a) and (b) corresponding to the two transmit antennas.}
\label{fig2}
\end{figure}

IAM preamble design method has drawn much attention due to its simplicity and efficiency \cite{Kofidis2013}. The IAM family that includes IAM-R,  IAM-imaginary (IAM-I), IAM-C, and extended IAM-C (E-IAM-C) were firstly proposed in the SISO system based on approximating the unknown intrinsic interference, where three columns of preamble symbols are placed in front of the data frame \cite{CE, C2, JinFeng, EIAMC}. Correspondingly, in the MIMO case, the IAM preambles were constructed by repeating the SISO preamble $N_{t}$ times for each transmit antenna, with some changed signs being inserted simultaneously to guarantee the orthogonality of different antennas \cite{Kofidis2011}. Taking $N_{t}=2$ as an example, Fig. \ref{fig2} depicts the IAM-C preamble at the SFB input with $M=8$ subcarriers.
According to (\ref{eq 2.10}), the demodulated symbols at times $n=1,3$ are obtained as \cite{Kofidis2013, Kofidis2011}
\begin{align}\label{eq 3.2}
\left[ ~\textbf{y}_{m,1}~~ \textbf{y}_{m,3}~ \right] = \textbf{H}_{m}
\left[ \begin{array}{cc}
c_{m,1}^{1}         &c_{m,3}^{1}           \\
c_{m,1}^{2}                 & c_{m,3}^{2}          \\
\end{array} \right]
+\left[ ~\bm{\eta}_{m,1}~~ \bm{\eta}_{m,3}~ \right].
\end{align}
Under the assumption that the intrinsic interference mostly originates from the first-order FT neighbors, the intrinsic interference to the pilots are computed approximately and hence the so-called pseudo-pilots are constructed for channel estimation. We can easily see that the pseudo-pilots satisfy $c_{m,1}^{1}=c_{m,3}^{1}=c_{m,1}^{2}=- c_{m,3}^{2}=c_{m}$, where $c_{m}$ can be calculated as in the SISO case. Substituting it into (\ref{eq 3.2}) we get
\begin{equation}\label{eq 3.3}
\left[ ~\textbf{y}_{m,1}~~ \textbf{y}_{m,3}~ \right]  = \textbf{H}_{m}c_{m}\textbf{A}_{2}+\left[ ~\bm{\eta}_{m,1}~~ \bm{\eta}_{m,3}~ \right],
\end{equation}
where $\textbf{A}_{2}$ is the orthogonal matrix
$$\textbf{A}_{2}=\left[ \begin{array}{cc}
1         &1          \\
1                & -1          \\
\end{array} \right].$$
Hence, the least squares estimate of the CFR matrix at the $m$th subcarrier can be obtained as
\begin{align}\label{eq 3.4}
\hat{\textbf{H}}_{m}&=\left[ ~\textbf{y}_{m,1}~~ \textbf{y}_{m,3}~ \right] \frac{1}{c_{m}}\textbf{A}_{2}^{-1}\nonumber\\
&=\textbf{H}_{m}+\frac{1}{2c_{m}}\left[ ~\bm{\eta}_{m,1}~~ \bm{\eta}_{m,3}~ \right]\textbf{A}_{2}.
\end{align}

In the IAM method, all subcarriers are used to directly estimate the corresponding CFR values. It is worth noting that the length of the IAM preamble overhead is $2N_{t}+1$ ($3N_{t}$ for E-IAM-C) columns, where $N_{t}$ columns are used for channel estimation. Therefore, the IAM method encounters a considerable loss in spectral efficiency, especially for a relatively large number of transmit antennas.

\subsection{FDM Method}
\begin{figure}[!h]
\centering
\includegraphics[scale=1.1]{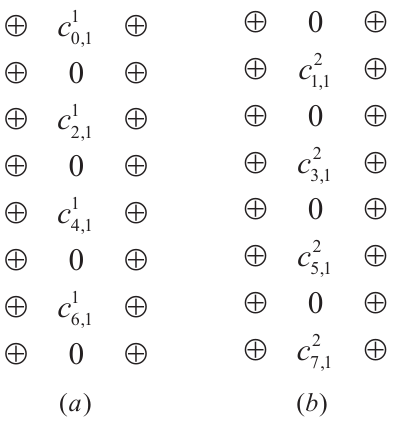}
\caption{Received pilots at the AFB output for the FDM method, with (a) and (b) corresponding to the two transmit antennas and $\oplus$ being the guard symbols.}
\label{fig3}
\end{figure}

To improve the spectral efficiency, the FDM preamble design method with the sparse type has been proposed by sharing the subcarriers among the antennas. This subsection provides a brief review of the FDM method presented in \cite{Taheri2015}. In the FDM method, the pilots are designed specially to guarantee that the pseudo-pilots at different subcarriers are of the frequency-division multiplexed type. As a consequence, the channels at the \emph{active} subcarriers can be estimated separately without inter-antenna interference. The frequency response values at the \emph{missing} subcarriers are then found via interpolation of neighboring subchannels. Fig. \ref{fig3} depicts an example for the case $N_{t}=2$ and $M=8$, in which $c_{m,n}^{t}$ is the pseudo-pilot (i.e., the ideal AFB output) corresponding to the $t$th transmit antenna.

As can be seen in Fig. \ref{fig3}, nonzero pseudo-pilots for channel estimation are distributed on odd/even subcarriers of the middle column, with the rest of them being nulled. According to (\ref{eq 2.8}), the corresponding demodulated symbols at the $r$th receive antenna can be expressed as
\begin{equation}\label{eq 3.5}
y_{m,1}^{r} =
\begin{cases}
H_{m}^{r,1}c_{m,1}^{1}+\eta_{m,1}^{r}, & m \ \rm {even},\\
H_{m}^{r,2}c_{m,1}^{2}+\eta_{m,1}^{r}, & m \ \rm {odd}.
\end{cases}
\end{equation}
Therefore, the least squares estimates of $H_{m}^{r,1}$ and $H_{m}^{r,2}$ are given by
\begin{equation}\label{eq 3.6}
\hat H_{m}^{r,1}=\frac{y_{m,1}^{r}}{c_{m,1}^{1}}=H_{m}^{r,1}+\frac{\eta_{m,1}^{r}}{c_{m,1}^{1}}, \  m\  \rm {even},
\end{equation}
\begin{equation}\label{eq 3.60}
\hat H_{m}^{r,2}=\frac{y_{m,1}^{r}}{c_{m,1}^{2}}=H_{m}^{r,2}+\frac{\eta_{m,1}^{r}}{c_{m,1}^{2}}, \  m\  \rm {odd},
\end{equation}
respectively. It is clear from  (\ref{eq 3.6}) and (\ref{eq 3.60}) that the CFR values at the active subcarriers are perfectly estimated irrespective of the additive noise. Accordingly, the CFR values at the missing subcarriers can be easily calculated by linear interpolation of neighboring subchannels.

 It is worth pointing out that the preamble overhead of the FDM method is fixed three columns of symbols, which shows overwhelming superiority in spectral efficiency compared to the IAM method. However, the preamble in the existing FDM method is obtained without theoretical analysis and there is no literature concerning the preamble optimization in terms of MSE performance currently. To enhance the estimation accuracy, it is necessary to design the corresponding optimal preamble for the FDM method.

\section{PROPOSED PREAMBLE OPTIMIZATION METHOD}
In this section, we propose a preamble optimization method for the FDM-structured preamble to improve the channel estimation performance in MIMO-OQAM/FBMC systems.

According to (\ref{eq 3.6}) and (\ref{eq 3.60}), it is clear that the FDM-structured preamble optimization results from: $(i)$ pseudo-pilots at the active subcarriers of maximum magnitudes to control the estimation noise power, and $(ii)$ pseudo-pilots at the missing subcarriers of zero values so that there is no interference on the corresponding points at the receiver side. To calculate the pseudo-pilot at each FT point, one needs to know the imaginary interference coefficients $\zeta_{m + p_{0},n + q_{0}}^{m,n}$ for the neighbors $(p_{0},q_{0})\in\Omega_{1}$ as in (\ref{eq 2.03}). It can be easily shown that the imaginary interference coefficients follow a specific pattern for any employed prototype filter \cite{Kofidis2013}. For odd $n$, the values of $\zeta_{m + p_{0},n + q_{0}}^{m,n}$ for the first-order FT neighbors are detailed in Table \ref{Table1}. The quantities $\delta, \beta, \gamma$ are constants for a certain system and can be computed previously. Note that, there might be some sign changes for other definitions of $\varphi_{0}$ \cite{Kofidis2011}.
\begin{table}[!h]
\renewcommand\arraystretch{1.6}
  \centering
  \caption{Values of the imaginary interference coefficients $\zeta_{m+p_{0},n+q_{0}}^{m,n}$}
\begin{tabular}{|c|c|c|c|}\hline
     $\zeta_{m+p_{0},n+q_{0}}^{m,n}$ & $q_{0}=-1$ & $q_{0}=0$ & $q_{0}=1$ \\ \hline
    $p_{0}=-1$ &$ j\delta$ & $j\beta$ &$ j\delta$ \\ \hline
   $ p_{0}=0$ & $-j\gamma$ &$ 1 $&$ j\gamma $\\ \hline
    $p_{0}=1$ & $ j\delta$ & $-j\beta $ & $ j\delta$ \\ \hline
  \end{tabular}\label{Table1}
\end{table}

\subsection{Preamble Optimization Problem}
Given the symmetric and mutual isolation nature of pseudo-pilots distribution shown in Fig. \ref{fig3}, we assume that preamble symbols are generated repeatedly on every $N_{t}$ adjacent subcarriers. For one specific antenna, the preamble matrix at the SFB input is given as
\begin{align}\label{eq 3.701}
 \textbf{A}=\left[ \begin{array}{ccc}
\vdots & \vdots & \vdots \\
a_{1} & a_{2} & a_{3} \\
a_{4} & a_{5} & a_{6} \\
\vdots & \vdots &\vdots \\
a_{3N_{t}-2} & a_{3N_{t}-1} & a_{3N_{t}} \\
\vdots & \vdots & \vdots \\
\end{array} \right],
\end{align}
where $a_{i},\ i=1,2,\cdots,3N_{t}$ represent the transmitted pilots, with $a_{2}$ being transmitted at the active subcarrier and $a_{3i-1},\ i=2,3,\cdots,N_{t}$ being transmitted at the missing subcarriers. The pseudo-pilots at corresponding points are denoted by $c_{3i-1},\ i=1,2,\cdots,N_{t}$, which can be calculated according to (\ref{eq 2.9}) and the imaginary interference coefficients in Table \ref{Table1}. More specifically, we can express the pseudo-pilot as $c_{3i-1}=a_{3i-1}+ja_{3i-1}^{(i)}$, where $ja_{3i-1}^{(i)}$ is the intrinsic interference from the neighboring FT points to $a_{3i-1}$. Analogously, the preambles transmitted at other antennas can be taken as the frequency shifted versions of those at this antenna.

In this paper, we formulate an optimization problem to determine the optimal preamble, where the objective is to minimize the channel estimation MSE, subject to a total
training energy constraint per antenna. According to  (\ref{eq 3.6}) and (\ref{eq 3.60}), the channel estimation MSE at the active subcarrier is given by
\begin{equation}\label{eq 3.702}
\textrm{MSE}=\mathbb{E} \left(|\hat{H}_{m}-H_{m}|^{2} \right)=\frac{\sigma^2}{|c_{2}|^2}.
\end{equation}
It is clear from (\ref{eq 3.702}) that the larger the magnitude of the pseudo-pilot $c_{2}$ is, the better the channel estimation will be. Then, we consider the training energy constraint. It should be noted that there is a difference between the energy input to the SFB and the energy of the modulated preamble at the SFB output due to the fact that the existence of complex-valued pilot symbols would destroy the real orthogonality of the OQAM/FBMC system. In our approach, more realistically, the transmit energy constraint of the preamble is translated at the SFB output as in \cite{Kofidis2017}. Thus, the preamble optimization problem is formulated as
\begin{subequations}\label{eq 3.703}
\begin{align}
\min\limits_{a_{1},a_{2},...,a_{3N_{t}}}~~&\textrm{MSE}=\sigma^2/{|c_{2}|^2},\label{eq 3.703a}\\
\text{subject to}~~~&c_{3i-1}=0,~i=2,3,\cdots,N_{t}\label{eq 3.703b}\\
~~~~~~&\textrm{P}_{out}\leq\varepsilon,\label{eq 3.703c}
\end{align}
\end{subequations}
where the constraint (\ref{eq 3.703b}) means the pseudo-pilots at the missing subcarriers are forced to zero and (\ref{eq 3.703c}) refers to the SFB output energy constraint (see Appendix A for the expression of $\textrm{P}_{out}$). In the following subsections, we will discuss the different solving processes for $N_{t}=2$ and $N_{t}>2$, respectively.

\subsection{Closed-form Solution for $N_{t}=2$}
For two transmit antennas, the preamble matrix at the SFB input is given as
\begin{align}\label{eq 3.7}
 \textbf{A}=\left[ \begin{array}{ccc}
\vdots & \vdots & \vdots \\
a_{1} & a_{2} & a_{3} \\
a_{4} & a_{5} & a_{6} \\
a_{1} & a_{2} & a_{3} \\
a_{4} & a_{5} & a_{6} \\
\vdots & \vdots & \vdots \\
\end{array} \right],
\end{align}
where $a_{2}$ is transmitted at the active subcarrier and $a_{5}$ is transmitted at the missing subcarrier. Let $\textbf{a}=\left[a_{1},a_{2},a_{3},a_{4},a_{5},a_{6}\right]^T\in\mathbb{C}^{6}$ and $\textbf{w}=\left[j\gamma,1,-j\gamma,-j2\delta,0,-j2\delta\right]^T$ denote the preamble vector and the interference weighting vector, respectively. Under the assumption of the interference being mostly contributed first-order neighboring FT points,  the energy of the modulated preamble at the SFB output becomes
\begin{equation}\label{eq 3.008}
\textrm{P}_{out}=\frac{M}{2}\textbf{a}^{H}\textbf{B}\textbf{a},
\end{equation}
where
$$\textbf{B}=\left(
                        \begin{array}{cccccc}
                          1 & j\gamma & 0 & 0 & -j2\delta & 0 \\
                          -j\gamma & 1 & j\gamma & j2\delta & 0 & j2\delta \\
                          0 & -j\gamma & 1 & 0 & -j2\delta & 0 \\
                          0 & -j2\delta & 0 & 1 &  j\gamma & 0 \\
                          j2\delta & 0 & j2\delta & -j\gamma & 1 & j\gamma \\
                          0 & -j2\delta & 0 & 0 & -j\gamma & 1 \\
                        \end{array}
                      \right)$$
 is a full-rank Hermitian matrix. Besides, according to (\ref{eq 2.9}), the pseudo-pilots at corresponding points are obtained as
\begin{equation}\label{eq 3.8}
\begin{cases}
c_{2}=a_{2}+j\left[(2a_{4}+2a_{6})\delta+(a_{3}-a_{1})\gamma \right]=\textbf{w}^H\textbf{a},\\
c_{5}=a_{5}+j\left[(2a_{1}+2a_{3})\delta+(a_{6}-a_{4})\gamma \right]=\textbf{w}^H\textbf{T}\textbf{a},
\end{cases}
\end{equation}
where
$$\textbf{T}=\left[ \begin{array}{cc}
\textbf{0}_{3}         &\textbf{I}_{3}           \\
\textbf{I}_{3}                & \textbf{0}_{3}           \\
\end{array} \right]$$
is a $6\times6$ transformation matrix. Resorting to matrix notation the optimization problem (\ref{eq 3.703}) is modified as
\begin{subequations}\label{eq 3.12}
\begin{align}
\min\limits_{\textbf{a}\in\mathbb{C}^{6}}~~~&-{|c_{2}|^2}=-|\textbf{w}^H\textbf{a}|^2,\label{eq 3.12a}\\
\text{subject to}~~~~&\textbf{w}^H\textbf{T}\textbf{a}=0,\label{eq 3.12b}\\
~~~~~~&\frac{M}{2}\textbf{a}^{H}\textbf{B}\textbf{a}\leq\varepsilon.\label{eq 3.12c}
\end{align}
\end{subequations}
Without loss of generality, we assume $\varepsilon=M/2$. Then, the above optimization problem (\ref{eq 3.12}) becomes
\begin{subequations}\label{eq 3.012}
\begin{align}
\min\limits_{\textbf{a}\in\mathbb{C}^{6}}~~~&-|\textbf{w}^H\textbf{a}|^2,\label{eq 3.012a}\\
\text{subject to}~~~~&\textbf{w}^H\textbf{T}\textbf{a}=0,\label{eq 3.012b}\\
~~~~~~&\textbf{a}^{H}\textbf{B}\textbf{a}\leq1.\label{eq 3.012c}
\end{align}
\end{subequations}
As stated and proved in Appendix B, an optimal closed-form solution to (\ref{eq 3.012}) is given by
\begin{equation}\label{eq 3.013}
\textbf{a}=\textbf{B}^{-1}\textbf{w}.
\end{equation}
This solution indicates that in the case of two transmit antenna, the optimal preamble depends only on the imaginary interference coefficients and can be pre-calculated based on the prototype filter $g[k]$ employed.

\subsection{Suboptimal Solution for $N_{t}>2$}
Given the symmetric and fast decay nature of the imaginary interference coefficients as in Table \ref{Table1}, the preamble optimization problem for any $N_{t}>2$ would be analogous. Therefore, we take $N_{t}=4$ as an example and give a general solving process. In this case, the preamble matrix of one antenna at the SFB input is given as
\begin{align}\label{eq 6.1}
 \textbf{A}=\left[ \begin{array}{ccc}
\vdots & \vdots & \vdots \\
a_{1} & a_{2} & a_{3} \\
a_{4} & a_{5} & a_{6} \\
a_{7} & a_{8} & a_{9} \\
a_{10} & a_{11} & a_{12} \\
\vdots & \vdots & \vdots \\
\end{array} \right],
\end{align}
where $a_{2}$ is transmitted at the active subcarrier and $a_{5},a_{8},a_{11}$ are transmitted at the missing subcarriers. Denoting by $\textbf{a}=\left[a_{1},a_{2},\ldots, a_{12}\right]^T\in\mathbb{C}^{12}$ the preamble vector, by $\textbf{w}=[-j\delta,-j\beta,-j\delta,j\gamma,1,-j\gamma,-j\delta,j\beta,-j\delta]^{T}$ the corresponding interference weighting vector, the energy of the modulated preamble at the SFB output becomes
\begin{equation}\label{eq 6.02}
\textrm{P}_{out}=\frac{M}{4}\textbf{a}^{H}\textbf{\={B}}\textbf{a},
\end{equation}
where $\textbf{\={B}}\in\mathbb{C}^{12\times12}$ can be pre-calculated analogously (the expression of $\textbf{\={B}}$ is omitted here for simplicity). Besides, the pseudo-pilots at corresponding points can be expressed as
\begin{equation}\label{eq 6.2}
\begin{cases}
c_{2}=\textbf{w}^{H}\textbf{T}_{0}\textbf{a},\\
c_{5}=\textbf{w}^{H}\textbf{T}_{1}\textbf{a},\\
c_{8}=\textbf{w}^{H}\textbf{T}_{2}\textbf{a},\\
c_{11}=\textbf{w}^{H}\textbf{T}_{3}\textbf{a},
\end{cases}
\end{equation}
where
$$\textbf{T}_{0}=\left[ \begin{array}{cccc}
\textbf{0}_{3}  &\textbf{0}_{3}  &\textbf{0}_{3}       &\textbf{I}_{3}           \\
\textbf{I}_{3}                & \textbf{0}_{3}    & \textbf{0}_{3} & \textbf{0}_{3}       \\
\textbf{0}_{3}  &\textbf{I}_{3}                & \textbf{0}_{3}    & \textbf{0}_{3}  \\
\end{array} \right],$$
$$\textbf{T}_{1}=\left[ \begin{array}{cccc}
\textbf{I}_{3}                & \textbf{0}_{3}    & \textbf{0}_{3} & \textbf{0}_{3}       \\
\textbf{0}_{3}  &\textbf{I}_{3}                & \textbf{0}_{3}    & \textbf{0}_{3}  \\
\textbf{0}_{3}  &\textbf{0}_{3} &\textbf{I}_{3}  &\textbf{0}_{3}                  \\
\end{array} \right],$$
$$\textbf{T}_{2}=\left[ \begin{array}{cccc}
\textbf{0}_{3}  &\textbf{I}_{3}                & \textbf{0}_{3}    & \textbf{0}_{3}  \\
\textbf{0}_{3}  &\textbf{0}_{3} &\textbf{I}_{3}  &\textbf{0}_{3}                  \\
\textbf{0}_{3}  &\textbf{0}_{3}  &\textbf{0}_{3}       &\textbf{I}_{3}           \\
\end{array} \right],$$
$$\textbf{T}_{3}=\left[ \begin{array}{cccc}
\textbf{0}_{3}  &\textbf{0}_{3} &\textbf{I}_{3}  &\textbf{0}_{3}                  \\
\textbf{0}_{3}  &\textbf{0}_{3}  &\textbf{0}_{3}       &\textbf{I}_{3}           \\
\textbf{I}_{3}                & \textbf{0}_{3}    & \textbf{0}_{3} & \textbf{0}_{3}       \\
\end{array} \right]$$
are both $9\times12$ transformation matrices. We assume the total training energy constraint $\varepsilon=M/4$. According to the above derivations, the original optimization problem (\ref{eq 3.703}) for $N_{t}=4$ is modified as
\begin{subequations}\label{eq 6.3}
\begin{align}
\min\limits_{\textbf{a}\in\mathbb{C}^{12}}~~&-|\textbf{w}^{H}\textbf{T}_{0}\textbf{a}|^2,\label{eq 6.3a}\\
\text{subject to}~~~&\textbf{w}^{H}\textbf{T}_{1}\textbf{a}=0,\label{eq 6.3b}\\
&\textbf{w}^{H}\textbf{T}_{2}\textbf{a}=0,\label{eq 6.3c}\\
&\textbf{w}^{H}\textbf{T}_{3}\textbf{a}=0,\label{eq 6.3d}\\
&\textbf{a}^{H}\textbf{\={B}}\textbf{a}\leq1.\label{eq 6.3e}
\end{align}
\end{subequations}
Recall that $|\textbf{w}^{H}\textbf{T}_{i}\textbf{a}|^{2}=\textbf{a}^{H}\left(\textbf{T}_{i}^{H}\textbf{w}\textbf{w}^{H}\textbf{T}_{i}\right)\textbf{a}$ and that $x=0\Leftrightarrow|x|^{2}=0$, for all $x\in\mathbb{C}$.
Hence, the above optimization problem (\ref{eq 6.3}) can be equivalently stated as
\begin{subequations}\label{eq 6.5}
\begin{align}
\min\limits_{\textbf{a}\in\mathbb{C}^{12}}~~~&-\textbf{a}^{H}\left(\textbf{T}_{0}^{H}\textbf{w}\textbf{w}^{H}\textbf{T}_{0}\right)\textbf{a},\label{eq 6.4a}\\
\text{subject to}~~~~&\textbf{a}^{H}\left(\textbf{T}_{1}^{H}\textbf{w}\textbf{w}^{H}\textbf{T}_{1}\right)\textbf{a}=0,\label{eq 6.4b}\\
~~~~~~&\textbf{a}^{H}\left(\textbf{T}_{2}^{H}\textbf{w}\textbf{w}^{H}\textbf{T}_{2}\right)\textbf{a}=0,\label{eq 6.4c}\\
~~~~~~&\textbf{a}^{H}\left(\textbf{T}_{3}^{H}\textbf{w}\textbf{w}^{H}\textbf{T}_{3}\right)\textbf{a}=0,\label{eq 6.4d}\\
~~~~~~&\textbf{a}^{H}\textbf{\={B}}\textbf{a}\leq1.\label{eq 6.4e}
\end{align}
\end{subequations}
It should be noted that the objective function (\ref{eq 6.4a}) and the constraints (\ref{eq 6.4b}), (\ref{eq 6.4c}), (\ref{eq 6.4d}), and (\ref{eq 6.4e}) are both quadratic. Therefore, the optimization problem (\ref{eq 6.5}) is a quadratically constrained quadratic program (QCQP). Solving the general case of QCQP is an NP-hard problem and therefore lacks computationally efficient solution.

Observe that the optimal solution is hard to obtain for the above nonconvex QCQP with complex variables. Therefore, in the following, we propose to convert the original nonconvex optimization into a convex optimization by relaxing the nonconvex constraint and obtain the corresponding suboptimal solution.

Denoting by $\textbf{C}_{i}=\textbf{T}_{i}^{H}\textbf{w}\textbf{w}^{H}\textbf{T}_{i},\ i=0,1,2,3$ the corresponding coefficient matrices, we firstly perform the trace transformation on the objective function and constraints as
\begin{equation}\label{eq 6.6}
\textbf{a}^{H}\textbf{C}_{i}\textbf{a} =\textrm{Tr}{ \left(\textbf{a}^{H}\textbf{C}_{i}\textbf{a}\right)}=\textrm{Tr}{ \left( \textbf{C}_{i}\textbf{a}\textbf{a}^{H}\right)},\ i=0,1,2,3
\end{equation}
\begin{equation}\label{eq 6.61}
\textbf{a}^{H}\textbf{\={B}}\textbf{a} =\textrm{Tr}{ \left(\textbf{a}^{H}\textbf{\={B}}\textbf{a}\right)}=\textrm{Tr}{ \left(\textbf{\={B}}\textbf{a}\textbf{a}^{H}\right)}.
\end{equation}
Correspondingly, the optimization problem (\ref{eq 6.5}) becomes
\begin{subequations}\label{eq 6.7}
\begin{align}
\min\limits_{\textbf{a}\in\mathbb{C}^{12}}~~~&-\textrm{Tr}{ \left(\textbf{C}_{0}\textbf{a}\textbf{a}^{H}\right)},\label{eq 6.7a}\\
\text{subject to}~~~~~&\textrm{Tr}{ \left(\textbf{C}_{i}\textbf{a}\textbf{a}^{H}\right)}=0,\ i=1,2,3\label{eq 6.7b}\\
~~~~~&\textrm{Tr}{\left( \textbf{\={B}}\textbf{a}\textbf{a}^{H}\right)}\leq1.\label{eq 6.7c}
\end{align}
\end{subequations}
In particular, both the objective function and constraints in (\ref{eq 6.7}) are linear in the matrix $ \textbf{a}\textbf{a}^{H}$. Therefore, we define a new variable $\textbf{X}=\textbf{a}\textbf{a}^{H}$, with $\textbf{X}$ being a rank one symmetric positive semidefinite matrix. Then, the optimization problem (\ref{eq 6.7}) is rewritten as
\begin{subequations}\label{eq 6.8}
\begin{align}
\min\limits_{\textbf{X}\in \mathbb{C}^{12\times12}}~~~&-\textrm{Tr}{\left(\textbf{C}_{0}\textbf{X}\right)},\label{eq 6.8a}\\
\text{subject to}~~~&\textrm{Tr}{\left(\textbf{C}_{i}\textbf{X}\right)}=0,\ i=1,2,3\label{eq 6.8b}\\
~~~~~&\textrm{Tr}{\left(\textbf{\={B}}\textbf{X}\right)}\leq1,\label{eq 6.8c} \\
~~~~~&\textbf{X}\succeq 0, \textrm{rank}(\textbf{X})=1,\label{eq 6.8d}
\end{align}
\end{subequations}
where $\succeq$ in (\ref{eq 6.8d}) indicates that \textbf{X} is symmetric positive semidefinite. Indeed, the only difficult constraint in (\ref{eq 6.8}) is the nonconvex rank constraint $\textrm{rank}(\textbf{X})=1$. By dropping the nonconvex rank-one constraint, (\ref{eq 6.8}) is relaxed as
\begin{subequations}\label{eq 6.9}
\begin{align}
\min\limits_{\textbf{X}\in \mathbb{C}^{12\times12}}~~~&-\textrm{Tr}{\left(\textbf{C}_{0}\textbf{X}\right)},\label{eq 6.9a}\\
\text{subject to}~~~&\textrm{Tr}{\left(\textbf{C}_{i}\textbf{X}\right)}=0,\ i=1,2,3\label{eq 6.9b}\\
~~~~~&\textrm{Tr}{\left(\textbf{\={B}}\textbf{X}\right)}\leq1,\label{eq 6.9c} \\
~~~~~&\textbf{X}\succeq 0.\label{eq 6.9d}
\end{align}
\end{subequations}
The problem (\ref{eq 6.9}) is convex and can be solved effectively using the convex optimization toolbox CVX \cite{CVX}. Denoting by $\textbf{\~{X}}$ a globally optimal solution to (\ref{eq 6.9}), the remaining problem is how to obtain the feasible solution $\textbf{\~{a}}$ to  (\ref{eq 6.7}) from $\textbf{\~{X}}$.

Resorting to the matrix eigendecomposition, the matrix $\textbf{\~{X}}$ is represented in terms of its eigenvalues and eigenvectors, i.e.,
\begin{equation}\label{eq 6.10}
\textbf{\~{X}}=\sum _{i=1}^{r}\lambda_{i}\textbf{u}_{i}\textbf{u}_{i}^{H},
\end{equation}
where $r$ is the rank of $\textbf{\~{X}}$ and $\lambda_{i}$ is a scalar, termed the eigenvalue corresponding to the eigenvector $\textbf{u}_{i}$. Note that, the best rank-one approximation $\textbf{\~{X}}_{m}$ to $\textbf{\~{X}}$ is given by
\begin{equation}\label{eq 6.11}
\textbf{\~{X}}_{m}=\lambda_{m}\textbf{u}_{m}\textbf{u}_{m}^{H},
\end{equation}
where $\lambda_{m}$ is the maximum eigenvalue of $\textbf{\~{X}}$. Thus, the candidate solution to (\ref{eq 6.7}) is accordingly obtained as $\textbf{\~{a}}=\sqrt{\lambda_{m}}\textbf{u}_{m}$. It should be noted that some additive map operations may also be needed to obtain a ``nearby'' feasible solution if $\sqrt{\lambda_{m}}\textbf{u}_{m}$ is not feasible, where the literature \cite{SDR} can be referred for a more detailed discussion.

\section{SIMULATION RESULTS}
In this section, simulations are conducted to evaluate the proposed preamble design method in MIMO-OQAM/FBMC systems. We compare the proposed method with the conventional IAM and  FDM methods under $2\times2$ and $4\times4$ MIMO configurations, respectively. The simulation parameters for the MIMO-OQAM/FBMC systems are as follows. There are $M=256$ subcarriers modulated by 4-QAM constellations and the subcarrier spacing is $10.94$ kHz.  Each data frame consists of 20 complex OQAM/FBMC symbols. The well-known PHYDYAS prototype filter with overlapping factor of $4$ is employed because of its strong frequency localization \cite{phydyas}. The multipath channel SUI-3 proposed by the IEEE 802.16 broadband wireless access working group is adopted for our simulations \cite{SUI-3}. Meanwhile, the MIMO-OQAM/FBMC receiver employs the LMMSE equalizer for  data detection.

With PHYDYAS filter, the imaginary interference coefficients are determined as in Table \ref{Table2} for odd time index $n$. Thus, the quantities in Section ~\Rmnum{4} are assigned as $\delta=0.2058, \beta=0.2393, \gamma=0.5644$ and the optimal preamble can be constituted accordingly. It is worth mentioning that the optimality of the proposed preamble is guaranteed on condition that the intrinsic imaginary interference only come from the first-order neighborhood. However, in the simulation experiments, the interference to the $(m,n)$ FT point from $(m\pm1,n\pm2)$ and $(m\pm1,n\pm3)$ is non-negligible as detailed in Table \ref{Table2}. In consequence, the residual interference from data symbols would degrade the channel estimation performance. Therefore, we also propose the extended versions of the optimal preamble by following the pilot symbols with one and two columns of zero symbols, which are defined as ``$G=1$" and ``$G=2$", respectively. For fair comparison, the preambles of all methods are normalized to equal power at the SFB output.
\begin{table*}[]
\renewcommand\arraystretch{1.2}
  \centering
  \caption{Values of the imaginary interference coefficients when PHYDYAS filter is used.}
\begin{tabular}{|c|c|c|c|c|c|c|c|}
\hline
     $\zeta_{m+p_{0},n+q_{0}}^{m,n}$ & $q_{0}=-3$ & $q_{0}=-2$ & $q_{0}=-1$ & $q_{0}=0$ & $q_{0}=1$& $q_{0}=2$& $q_{0}=3$ \\ \hline
    $p_{0}=-1$  & $ 0.0429j$ &$0.125j$&$0.2058j$ &$0.2393j$ &$0.2058j$ &$0.125j$&$0.0429j$\\
    \hline
   $ p_{0}=0$ & $-0.0668j$ & $0$ & $-0.5644j$ & $1$ & $0.5644j$ & $0$ & $0.0668j$  \\
    \hline
    $p_{0}=1$ & $0.0429j$ & $-0.125j$ & $0.2058j$ & $-0.2393j$ & $0.2058j$ & $-0.125j$ & $0.0429j$  \\
    \hline
  \end{tabular}\label{Table2}
\end{table*}

\begin{table*}[]
  \centering
\caption{Preamble overhead comparison for channel estimation in MIMO-OQAM/FBMC systems.}\label{Table3}
\begin{tabular}{|c|c|c|c|c|c|c|c|}
  \hline
  \multirow{3}{*}{Transmit antennas} & \multicolumn{7}{c|}{Preamble overhead}\\
  \cline{2-8}
  &  \multirow{2}{*}{IAM-R} & \multirow{2}{*}{ IAM-C}  &  \multirow{2}{*}{E-IAM-C } &  \multirow{2}{*}{\tabincell{c}{Conventional \\  FDM [24]} }& \multicolumn{3}{c|}{Proposed method}  \\
  \cline{6-8}
  & & & & & without null & $G=1$ & $G=2$ \\
\hline
$N_{t}=2$ &  $5$ & $5$ & $6$ & $3$& $3$ & $4$ & $5$ \\\hline
$N_{t}=4$ & $9$ & 9 & $12$ &$3$& $3$ & $4$ & $5$  \\\hline
$N_{t}$ & $2N_{t}+1$& $2N_{t}+1$  & $3N_{t}$ & $3$& $3$ & $4$ & $5$   \\
\hline
\end{tabular}
\end{table*}

The preamble overhead comparison among different preamble design methods is summarized in Tables \ref{Table3}, which shows great superiority of our proposed method compared to the IAM variants in terms of spectral efficiency. Then, the comparison of peak-to-average power ratio (PAPR) performance is conducted among different methods. It is not surprising that the periodic preamble structure of the proposed method and the three IAM variants, namely IAM-R, IAM-C, and E-IAM-C, would result in high PAPR signal. Nonetheless, the proposed preamble still yields the best PAPR characteristics at the SFB output (PAPR$\approx$22.8 dB) compared to that of the IAM-R ($\approx$24.4 dB), IAM-C ($\approx$25.6 dB), and E-IAM-C ($\approx$26.5 dB) preambles. In the following we will compare the MSE and BER performances of different preamble design methods under $2\times2$ and $4\times4$ MIMO configurations, respectively.

\begin{figure}[htb]
\centering
\includegraphics[scale=0.36]{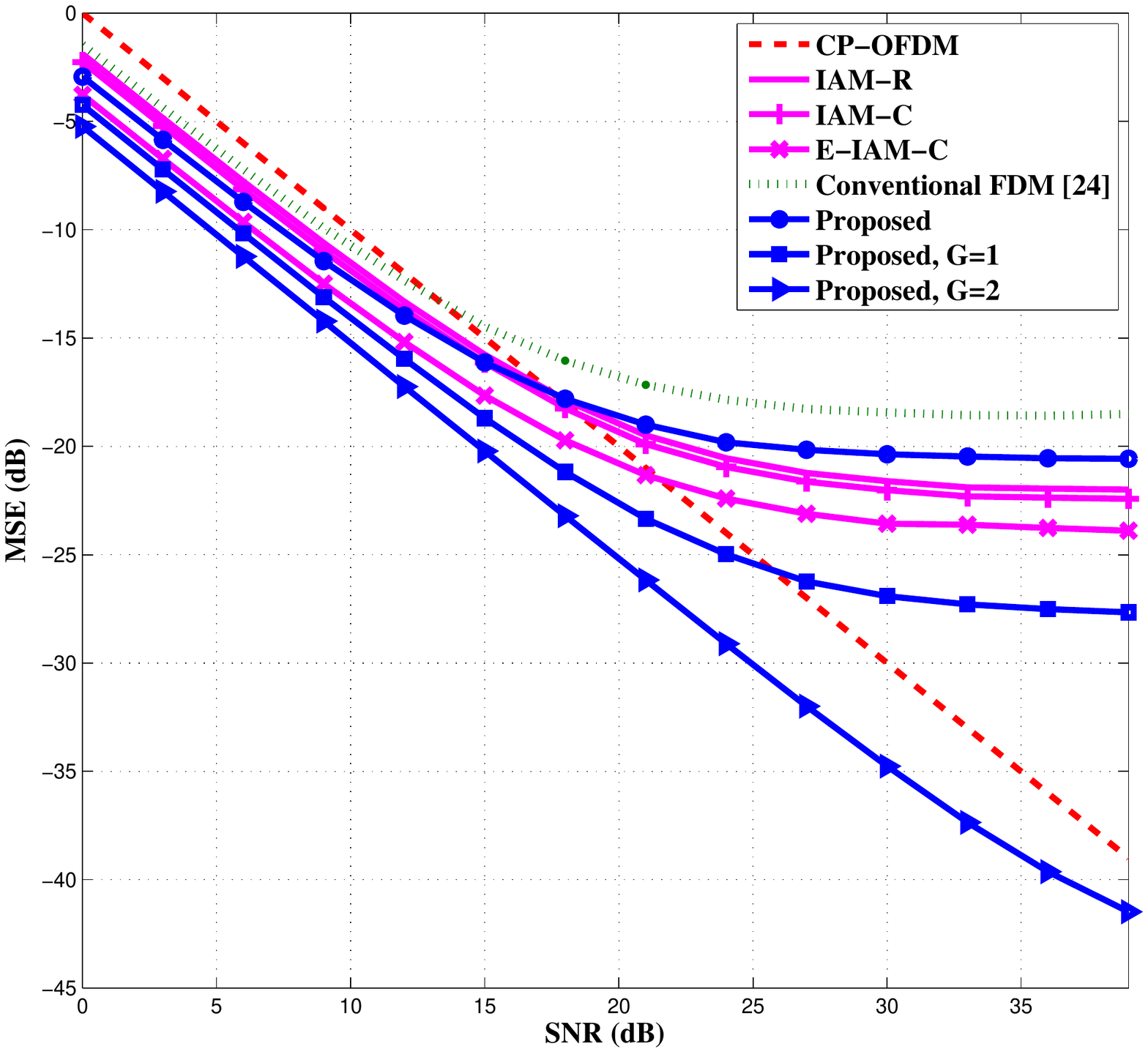}
\caption{Comparison of MSE performances among different preamble design methods for a 2 $\times$ 2 MIMO-OQAM/FBMC system.}
\label{fig5}
\end{figure}

\begin{figure}[htb]
\centering
\includegraphics[scale=0.36]{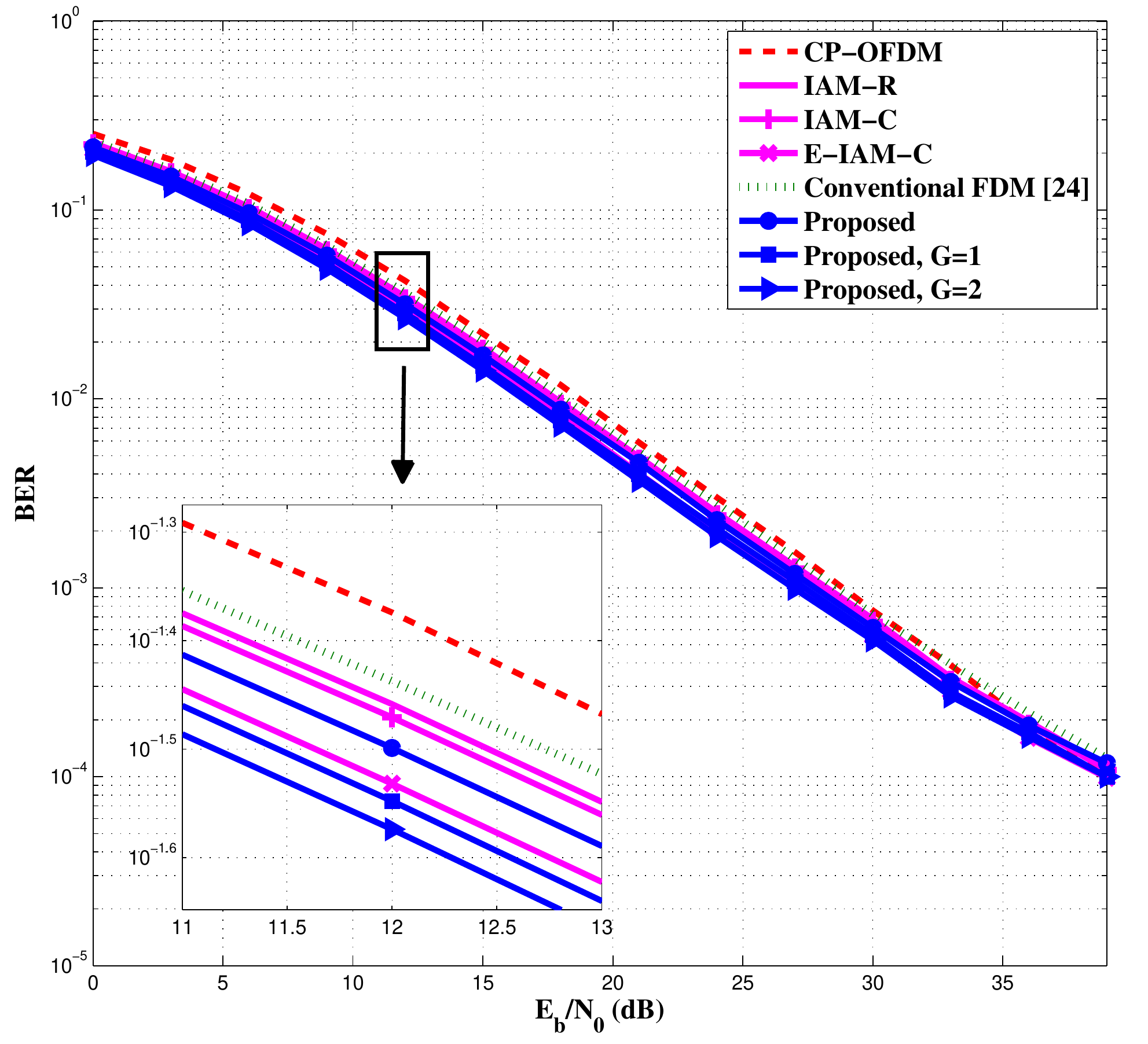}
\caption{Comparison of BER performances among different preamble design methods for a 2 $\times$ 2 MIMO-OQAM/FBMC system.}
\label{fig6}
\end{figure}

The case of two transmit antennas is considered firstly. Fig. \ref{fig5} and Fig. \ref{fig6} show the MSE and BER performances of different preamble design methods for a 2 $\times$ 2 MIMO-OQAM/FBMC system, respectively. For the sake of completeness, the performance curves of the CP-OFDM system are also presented with the CP length being chosen as $\frac{M}{8}$ according to the LTE specification \cite{OFDMop}. Since the residual interference is non-negligible as in Table \ref{Table2}, the pseudo-pilots in (\ref{eq 3.2}) differ from each other and the correct pseudo-pilots are used in the simulations. Besides, the intrinsic real-valued interferences from the imaginary-valued pilot symbols to data symbols are also considered for accurate data detection.

As observed in Fig. \ref{fig5}, the proposed method exhibits a significant performance gain over the conventional FDM method for all SNR regions and all methods perform better than CP-OFDM at low SNRs. The well-known error floor of the IAM variants and the proposed method at high SNRs is also observed due to the unavoidable intrinsic interference. Besides, the proposed method outperforms the IAM-R and IAM-C methods at low to medium SNRs. The reason is that the proposed method contributes to the maximum magnitudes of pseudo-pilots at the active subcarriers, which weakens the effect of noise. However, in the case of relatively higher SNRs, the noise becomes nonsignificant and the interference pre-cancellation errors (including the residual intrinsic interference which comes from the invalidation of model (2) and the error brought by linear interpolation of neighboring subchannels) in the proposed method prevail, which accounts for the reduced decrease speed of the proposed curve. Nevertheless, we can see that the proposed preamble with $G=1$ achieves channel estimation MSE smaller than the IAM variants and the error floor of the proposed method is almost completely removed if two columns of zero symbols are placed immediately behind the pilots. The corresponding uncoded BER performance comparison and a zoomed-in version of part of BER curves are presented in Fig. \ref{fig6}. As observed, our proposed method achieves BER performance comparable to the IAM variants.

To verify the theoretical result in subsection IV-C, the simulation is also conducted for the MIMO $4\times4$ configuration, with the simulation parameters being set up similarly. For the proposed method, the approximate optimal solutions of $a_{1}, a_{2}, ... , a_{12}$ are calculated as $0, 0.9207, 0, 0.1952, 0, -0.1952, 0, 0, 0, -0.1952, 0, 0.1952$ according to the solving
process in subsection IV-C. Hence, the preambles transmitted at each of the transmit antennas are obtained by the corresponding repetition and frequency shift operations. Besides, the IAM variants are also tested for the sake of the comparison, where the changed signs in (\ref{eq 3.3}) are selected as a Hadamard matrix of order four \cite{Kofidis2011}.
\begin{figure}[htb]
\centering
\includegraphics[scale=0.36]{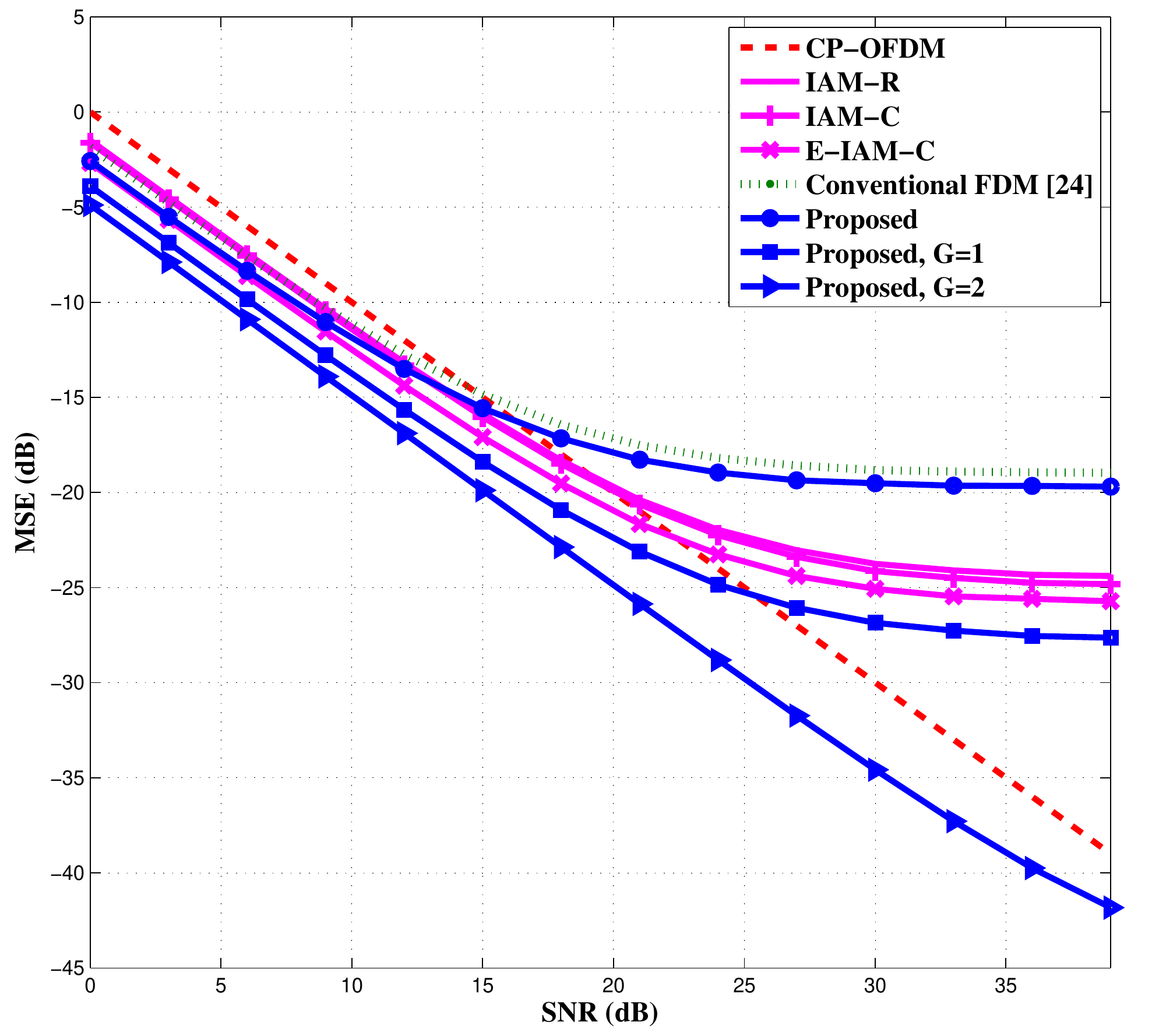}
\caption{Comparison of MSE performances among different preamble design methods for a 4 $\times$ 4 MIMO-OQAM/FBMC system.}
\label{fig7}
\end{figure}

\begin{figure}[htb]
\centering
\includegraphics[scale=0.36]{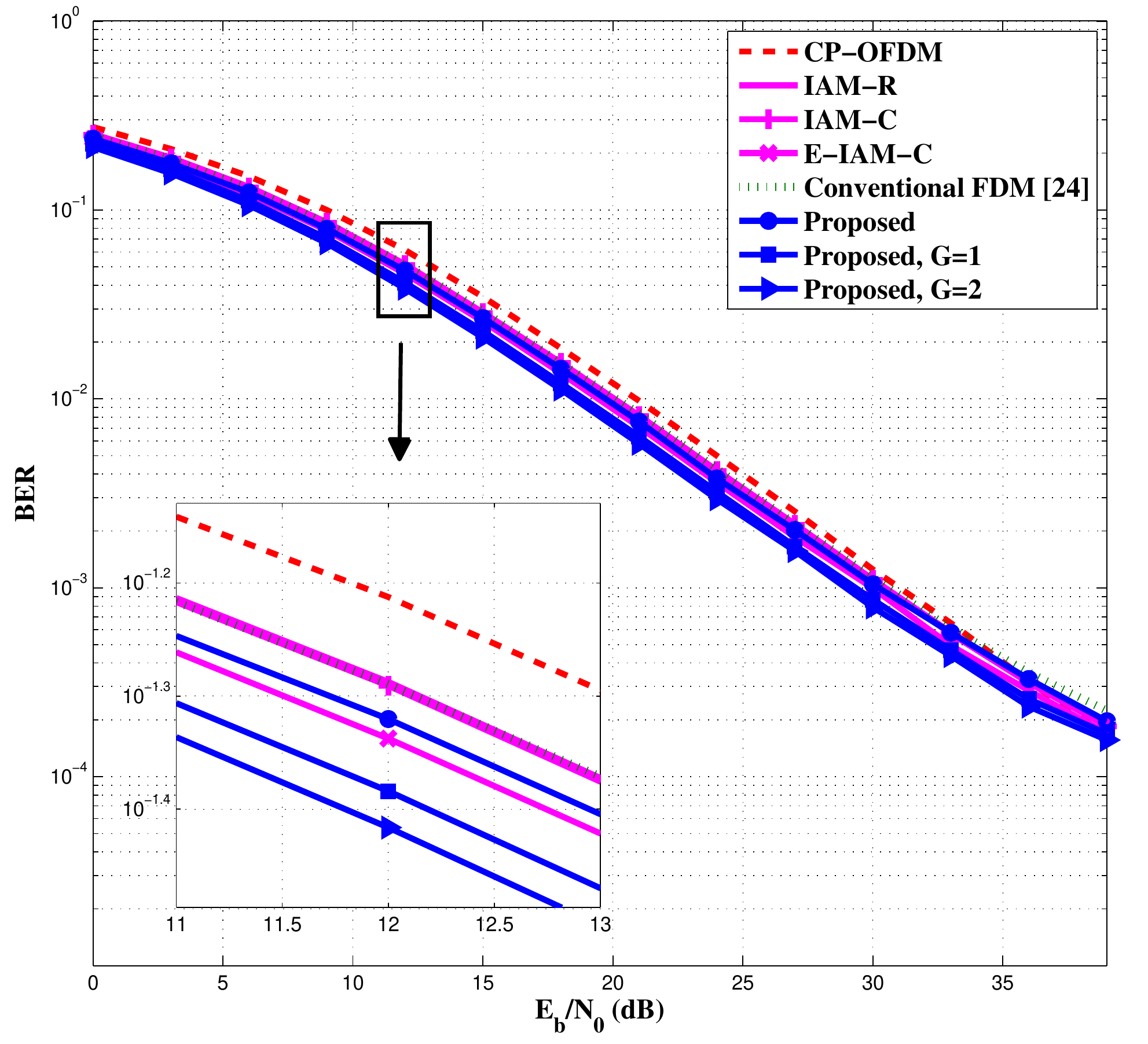}
\caption{Comparison of BER performances among different preamble design methods for a 4 $\times$ 4 MIMO-OQAM/FBMC system.}
\label{fig8}
\end{figure}

In the 4 $\times$ 4 MIMO-OQAM/FBMC system, the MSE and uncoded BER performances of different preamble design methods are depicted in Fig. \ref{fig7} and \ref{fig8}, respectively. From the simulation results, the proposed method with the optimized preamble outperforms the conventional FDM method at all SNR regimes in terms of MSE and BER performances. Besides, the proposed method performs slightly better than the IAM-R and IAM-C methods at low to medium SNRs, which validates the efficiency of the proposed method.
However, since three fourths of subchannels are obtained by the linear interpolation of neighboring subchannels, the channel estimation performance of the proposed method is worse than the E-IAM-C method due to the interference pre-cancellation errors. Nevertheless, as is shown in Table \ref{Table3}, the preamble overhead of the E-IAM-C method at each transmit antenna is twelve columns of symbols, while only three columns of symbols are needed with the proposed method. Besides, the proposed preamble with zero symbols being placed behind could significantly improve the channel estimation MSE and BER performances as in the two transmit antennas case. Thus, the proposed method provides a satisfactory compromise between channel estimation performance and preamble overhead to systems with more than two transmit antennas.

\section{CONCLUSIONS}
In this paper, we formulated an optimization problem to determine the frequency division multiplexed preambles in MIMO-OQAM/FBMC systems and gave the corresponding solutions for two transmit antennas and more than two transmit antennas, respectively. In the case of two transmit antennas, we found the relationship between preambles and the intrinsic interference from neighboring symbols to minimize the MSE of the channel estimation, and derived the optimal closed-form solution. In the case of more than two transmit antennas, we presented a general solving process to obtain the suboptimal solution by relaxing the nonconvex constraint. The MSE and BER performances of different preamble design methods were compared through simulations under 2$\times$2 and 4$\times$4 MIMO configurations, respectively. Simulation results have demonstrated that the proposed method with the optimized preamble shown great superiority compared with the conventional FDM method. Besides, a better channel estimation performance can be achieved than the IAM-R and IAM-C method at low to medium SNR regimes. Although a performance worse than the E-IAM-C method was experienced, the proposed method only requires fixed three columns of preamble symbols, resulting in considerable improvement in spectral efficiency compared with the IAM variants.

\begin{appendices}
      \section{the energy of the modulated preamble at the SFB output}
According to (\ref{eq 2.01}), the energy of the modulated preamble at the SFB output can be derived as
\begin{align}\label{eq 8.1}
\textrm{P}_{out} &=\sum_{k=-\infty}^{+\infty}s[k]s[k]^{\ast}   \nonumber\\
&=\sum_{k=-\infty}^{+\infty}\bigg\{\bigg[\sum_{m_{1}=0}^{M-1}\sum_{n_{1}=0}^{2}a_{m_{1},n_{1}}g_{m_{1},n_{1}}[k]\bigg] \nonumber\\ &\phantom{=}\qquad\qquad\times\bigg[\sum_{m_{2}=0}^{M-1}\sum_{n_{2}=0}^{2}a_{m_{2},n_{2}}^{\ast}g^{\ast}_{m_{2},n_{2}}[k]\bigg]\bigg\}\nonumber\\
&=\sum_{m_{1}=0}^{M-1}\sum_{m_{2}=0}^{M-1}\sum_{n_{1}=0}^{2}\sum_{n_{2}=0}^{2}\bigg\{a_{m_{1},n_{1}}a_{m_{2},n_{2}}^{\ast}\nonumber\\
&\phantom{=}\qquad\qquad\times\sum_{k=-\infty}^{+\infty}g_{m_{1},n_{1}}[k]g^{\ast}_{m_{2},n_{2}}[k]\bigg\}\nonumber\\
&=\sum_{m_{1}=0}^{M-1}\sum_{m_{2}=0}^{M-1}\sum_{n_{1}=0}^{2}\sum_{n_{2}=0}^{2}a_{m_{1},n_{1}}a_{m_{2},n_{2}}^{\ast}\zeta_{m_{1},n_{1}}^{m_{2},n_{2}}.
\end{align}
Let us denote by $\textbf{d}_{n}=\left[a_{0,n},a_{1,n},a_{2,n},\ldots,a_{M-1,n}\right]^T, n=0,1,2$ the preamble vectors. Define $\textbf{V}_{p,q}$ as the transmultiplexer response matrix of size $M\times M$ whose elements are given by $\left[\textbf{V}_{p,q}\right]_{i,j}=\zeta_{j,q}^{i,p}$ for $i\in\{0,1,\ldots,M-1\}$ and $j\in\{0,1,\ldots,M-1\}$. Then, we can easily rewrite (\ref{eq 8.1}) as
\begin{align}\label{eq 8.2}
\textrm{P}_{out}&=\left[\begin{array}{ccc}
                   \textbf{d}_{0}^{H} & \textbf{d}_{1}^{H} & \textbf{d}_{2}^{H}
                 \end{array}\right]\underbrace{\left[\begin{array}{ccc}
                                     \textbf{V}_{0,0} & \textbf{V}_{0,1} & \textbf{V}_{0,2} \\
                                     \textbf{V}_{1,0} & \textbf{V}_{1,1} & \textbf{V}_{1,2} \\
                                     \textbf{V}_{2,0} & \textbf{V}_{2,1} & \textbf{V}_{2,2}
                                   \end{array}\right]}_\textbf{V}\underbrace{\left[\begin{array}{c}
                                                                  \textbf{d}_{0} \\
                                                                  \textbf{d}_{1} \\
                                                                  \textbf{d}_{2}
                                                                \end{array}\right]}_\textbf{d}\nonumber\\
&=\textbf{d}^{H}\textbf{V}\textbf{d},
\end{align}
where $\textbf{d}\in\mathbb{C}^{3M}$ is the combined preamble vector and $\textbf{V}\in\mathbb{C}^{3M\times 3M}$ is the transmultiplexer response matrix. It is noteworthy that $\textbf{V}$ can be pre-calculated based on the prototype filter $g[k]$ employed.

      \section{ closed-form solution for the problem (\ref{eq 3.012})}
It is clear that the matrix $\textbf{B}$ in (\ref{eq 3.008}) is unitarily equivalent to a diagonal matrix due to its Hermitian property. Let us denote by $\textbf{B}=\textbf{U}\bm{\Sigma}\textbf{U}^{H}$, with $\textbf{U}$ unitary matrix and $\bm{\Sigma}$ real diagonal matrix, the eigenvalue decomposition of the Hermitian matrix $\textbf{B}$. Define $\textbf{\~{a}}=\bm{\Sigma}^{1/2}\textbf{U}^{H}\textbf{a}$ as the new variable. Then, the optimization problem (\ref{eq 3.012}) can be equivalently restated as
\begin{subequations}\label{eq 9.1}
\begin{align}
\min\limits_{\textbf{\~{a}}\in\mathbb{C}^{6}}~~~&-|\textbf{w}^H\textbf{U}\bm{\Sigma}^{-1/2}\textbf{\~{a}}|^2,\label{eq 9.1a}\\
\text{subject to}~~~~&\textbf{w}^H\textbf{T}\textbf{U}\bm{\Sigma}^{-1/2}\textbf{\~{a}}=0,\label{eq 9.1b}\\
~~~~~~&\textbf{\~{a}}^{H}\textbf{\~{a}}\leq1.\label{eq 9.1c}
\end{align}
\end{subequations}
We may as well drop the constraint (\ref{eq 9.1b}) to obtain the following relaxed version of (\ref{eq 9.1})
\begin{subequations}\label{eq 9.2}
\begin{align}
\min\limits_{\textbf{\~{a}}\in\mathbb{C}^{6}}~~~&-|\bm{\kappa}^H\textbf{\~{a}}|^2,\label{eq 9.2a}\\
\text{subject to}~~~~&\textbf{\~{a}}^{H}\textbf{\~{a}}\leq1,\label{eq 9.2b}
\end{align}
\end{subequations}
where $\bm{\kappa}=\bm{\Sigma}^{-1/2}\textbf{U}^H\textbf{w}\in\mathbb{C}^{6}$. Based on the concepts and properties of  Hermitian inner product in linear algebra, it is clear that the objective function in (\ref{eq 9.2}) can be minimized on condition that $\textbf{\~{a}}$ is collinear with $\bm{\kappa}$, which means all components of $\textbf{\~{a}}$ are in the same ratio to the corresponding components of $\bm{\kappa}$. Thus, the optimal solution to (\ref{eq 9.2}) is
\begin{equation}\label{eq 9.3}
\textbf{\~{a}}=\bm{\kappa}=\bm{\Sigma}^{-1/2}\textbf{U}^H\textbf{w}.
\end{equation}
It can be readily checked that the solution (\ref{eq 9.3}) also satisfies (\ref{eq 9.1b}), which means $\bm{\kappa}$ is a feasible solution to (\ref{eq 9.1}). Note that, the optimal objective value of (\ref{eq 9.2}) is definitely not greater than that of (\ref{eq 9.1}). Therefore, we could conclude that $\bm{\kappa}$ yields the best objective among all vectors that satisfy the constraints in the problem (\ref{eq 9.1}). Hence, $\bm{\kappa}$ is also the optimal solution to (\ref{eq 9.1}) and the optimal solution to the original problem (\ref{eq 3.012}) is given by
\begin{equation}\label{eq 9.4}
\textbf{a}=\textbf{U}\bm{\Sigma}^{-1/2}\textbf{\~{a}}=\textbf{U}\bm{\Sigma}^{-1}\textbf{U}^{H}\textbf{w}=\textbf{B}^{-1}\textbf{w}.
\end{equation}
This solution produces closed-form expression for the optimal preamble in the case of two transmit antennas, which depends only on the employed prototype filter.
  \end{appendices}

\end{document}